\documentclass[fleqn,usenatbib]{mnras}
\usepackage{newtxtext,newtxmath}
\usepackage[T1]{fontenc}
\DeclareRobustCommand{\VAN}[3]{#2}
\let\VANthebibliography\thebibliography
\def\thebibliography{\DeclareRobustCommand{\VAN}[3]{##3}\VANthebibliography}

\usepackage{graphicx}	% Including figure files
\usepackage{amsmath}	% Advanced maths commands
\usepackage{xspace}

\newcommand{\obj}{H1429-0028\xspace}

\newcommand{\hi}{H\,\textsc{i}\xspace}
\newcommand{\ci}{[C\,\textsc{i}]\xspace}

\newcommand{\kmps}{km\,s$^{-1}$}
\title[MeerKAT discovery of $z=1$ gigamaser]{MeerKAT discovery of a high-redshift strongly-lensed hydroxyl gigamaser}
\author[Manamela et al.]{Thato E. Manamela$^{1}$, Roger P. Deane$^{1,2,3}$, Tariq Blecher$^{4,5}$, Ian Heywood$^{6,7,4,5}$, Athol J. Kemball$^{8}$, \newauthor Danail Obreschkow$^{9,10}$   \\
$^1$ Department of Physics, University of Pretoria, Hatfield, Pretoria, 0028, South Africa \\
$^{2}$Inter-University Institute for Data Intensive Astronomy, Department of Astronomy, University of Cape Town, Cape Town, South Africa \\
$^3$ Wits Centre for Astrophysics, University of the Witwatersrand, 1 Jan Smuts Avenue, 2000, Johannesburg, South Africa\\
$^{4}$Department of Physics and Electronics, Rhodes University, PO Box 94, Makhanda, 6140, South Africa \\
$^{5}$South African Radio Astronomy Observatory, Liesbeek House, River Park, Gloucester Road, Mowbray, 7700, South Africa \\
$^{6}$SKA Observatory, Jodrell Bank, Lower Withington, Macclesfield, SK11 9FT, UK \\
$^{7}$Astrophysics, Department of Physics, University of Oxford, Keble Road, Oxford, OX1 3RH, UK \\
$^{8}$ Department of Astronomy, University of Illinois at Urbana-Champaign 1002 W. Green Street, Urbana, IL 61801, USA \\
$^9$ International Centre for Radio Astronomy Research (ICRAR), M468, University of Western Australia, WA 6009, Australia \\
$^{10}$ ARC Centre of Excellence for All Sky Astrophysics in 3 Dimensions (ASTRO 3D)}
\date{Accepted 2026 February 09. Received 2026 February 09; in original form 2025 December 05}
\begin{document}
\label{firstpage}
\pagerange{\pageref{firstpage}--\pageref{lastpage}}
\maketitle

\begin{abstract}
\noindent At low redshifts, hydroxyl megamasers (OHMs) have been shown to trace galaxy mergers, obscured starbursts, high molecular gas densities, and candidate dual supermassive black hole systems. Given this astrophysical utility, exploring these sources at larger cosmological look-back times is therefore of key interest. While previous OHM surveys have been limited to redshifts of $z \lesssim 0.25$, the ability to expand the OHM frontier is significantly enhanced with new high-sensitivity radio facilities such as MeerKAT. In this Letter, we report the discovery of an OHM in the gravitational lens system HATLAS\,J142935.3-002836 at $z = 1.027$, the most distant OHM source yet detected. The spectrum has blended 1667 and 1665 MHz emission and exhibits a highly complex profile, with spectral components ranging in widths of $<8$~km\,s$^{-1}$ to $\sim300$~km\,s$^{-1}$. The integrated (magnification uncorrected) luminosity of log($L_{\rm OH}~/~L_{\odot}$) = 5.51~$\pm$~0.67 makes this the most apparently luminous OHM known to date. In the same wide-band dataset, we have also detected a previously unknown \hi absorption line. The signal-to-noise ratio of over 150 with just a 4.7~h observation highlights the potential that MeerKAT and the future Square Kilometre Array mid-frequency array offer to explore the high-redshift OHM universe.
\end{abstract}


%%%%%%%%%%%%%%%%%%%%%%%%%%%%%%%%%%%%%%%%%%%%%%%%%%

%%%%%%%%%%%%%%%%% BODY OF PAPER %%%%%%%%%%%%%%%%%%

\section{Introduction}
\label{sec:Intro}
Hydroxyl megamasers (OHM) are luminous, extragalactic spectral line sources. As in their Galactic counterparts, the emission in OHMs is dominated by the maser lines at 1665 and 1667\,MHz, with much weaker satellite lines at 1612 and 1720\,MHz. 
Since OHMs require luminous far infrared (IR) radiation to sustain the population inversion necessary for stimulated emission \citep{lo2005mega,lockett2008effect}, they are typically found in the nuclear regions of luminous and ultra-luminous infrared galaxies (LIRGs and ULIRGs), many of which are also major merger systems \citep{sales2015, Hekatelyne2018a, sales2019gemini}. Indeed, the integrated OH line luminosity of OHMs is strongly correlated with the far-infrared (FIR) luminosity of the host galaxy over several orders of magnitude, following a super-linear power law, $L_{\text{OH}} \propto \left(L_{\text{FIR}}\right)^{1.2}$ \citep{baan1992a,darling2002search,Glowacki,zhang2024fashi}. OHMs are compact sources, with sizes of the order $\sim10-100$\,pc revealed by high-resolution radio imaging \citep{pihlstrom2001evn,rovilos2003continuum,lo2005mega}. However, unlike Galactic OH masers, the emission line at 1667\,MHz is more luminous than the 1665\,MHz emission line, with a typical ratio of 9:5 in local thermodynamic equilibrium \citep[][and references therein]{lo2005mega}. 
Due to Doppler broadening in massive, often merging galaxies, OHMs typically have large line widths of order $\sim$100--1000\,\kmps \citep{pihlstrom2001evn,darling2005oh}. 

The basic OH masering mechanism is outlined in \citet{Baan1989}, which requires a combination of (i) a background radio source producing the seed photon field, (ii) a foreground far-infrared field of pumping photons with an appropriate spectral energy distribution to enable population inversion of a large OH reservoir; and (iii) a sufficiently high column density of velocity-coherent gas aligned with both the radio and far-infrared (FIR) emission components to enable OH masering amplification. The result is a convolution of the background radio components, the foreground FIR components that drive the pumping mechanism, and the velocity-coherent gas. Some of the most detailed investigations of specific examples of this picture include \citet{Baan2023} using high-resolution eMERLIN/EVN observations of the prototypical nearby maser system Arp\,220. Since most OHMs are merger systems with high FIR luminosity, their space density is expected to increase dramatically in the high-redshift universe.

There has been a resurgence of interest in OHM at intermediate to high redshift as the Square Kilometre Array (SKA) and its precursors/pathfinders dramatically improve detection prospects at cosmological distances \citep{hess2021apertif,Roberts2021,Glowacki,jarvis2023discovery,Button2024}. This boost in sensitivity and the accessible cosmological volume enables high-redshift OHM surveys to take advantage of strong gravitational lensing, further enhancing prospects for high-z OHM surveys. This approach has been demonstrated as a route to explore the OHM universe to redshifts comparable to other important spectral lines that stem from the interstellar medium of galaxies \citep{Button2024}. 

We have designed several MeerKAT surveys to explore this new OHM observational frontier, including the {\sl Herschel}-discovered HATLAS\,J142935.3-002836 system (\obj~hereafter), which is a strongly lensed major merger system at $z$ = 1.027. It was identified using a FIR/mm selection technique that selects highly magnified, dust-rich starforming galaxies \citep{Blain1996,Negrello2007,negrello2010detection,Hezaveh2011,Vieira2013}. Following its discovery, a suite of deep, high-resolution multi-wavelength observations ensued, including radio \citep[Jansky Very Large Array;][]{perley2011expanded}, millimetre \citep[Atacama Large Millimetre/submillimetre Array;][]{alma}, and in the near-infrared \citep[{\sl HST}/Keck; see][]{messias2014herschel, Calanog2014}. \citet{messias2014herschel} conducted an extensive analysis of the gas and dust properties, dynamical data, and derived galaxy parameters through SED modelling of \obj. They discovered a massive reservoir of molecular gas detected through CO, CS, and \ci molecular and atomic lines, coupled with a high FIR-derived star formation rate of SFR $\sim$ 394$\pm$90~M$_{\odot}$~yr$^{-1}$. The high-resolution imaging and millimetre/optical spectroscopy contributed to a well-constrained lens model with integrated magnifications of \textbf{$\mu$} = 8 – 10 at millimetre to near-infrared (NIR) wavelengths. The total stellar and interstellar mass of the system is estimated as $M_{\star}$ = 1.3$\pm0.6 \times$ 10$^{11}$~M$_{\odot}$ and $M_{\textmd{ISM}}$ = 4.6~$\pm~1.7 \times$~10$^{10}$~M$_{\odot}$, respectively. The foreground lens is a disk galaxy viewed edge-on at a redshift of $z = 0.218$, with an almost complete $\theta_{\rm E} \sim$ 0.7~arcsec Einstein ring (see Fig.~\ref{Fig:lensmodel}). As described in the lens model analyses \citep{Calanog2014,messias2014herschel}, the lensed object in \obj~is composed of two NIR components with an implied stellar mass ratio of approximately 1:3. These two NIR components combined with the large cold molecular gas velocity widths and high SFR argue strongly that \obj~is an ongoing major merger system. 

In this Letter, we present the detection of an OHM source in \obj, along with corresponding \hi absorption. This is the highest redshift OHM system detected to date and the wealth of multi-wavelength data, combined with its strongly-lensed, complex spectral profile, makes it a powerful astrophysical laboratory. It previews what will be possible with much larger samples at earlier cosmological epochs in the SKA era. In this work, we adopt $\Lambda$CDM cosmology parameters of $\Omega_{\rm m}$~=~0.3, $\Omega_{\Lambda}$~=~0.7, and $H_0$~=~70~km~s$^{-1}$~Mpc$^{-1}$.

\section{Observations and Data Processing}\label{sec:Observations-Dataprocessing-calibration}

The observations of \obj~were conducted with the MeerKAT telescope \citep{2009IEEEP..97.1522J, 2016mks..confE...1J}, from April 13th to 16th, 2021 with 62 of the 64 antennas (proposal ID: SCI-20210212-RD-01) in the UHF band (544 - 1088~MHz). We used 8-s correlator integration time and the 32k  configuration of the UHF-band , yielding 32,768 channels and a channel width of 16.602~kHz (6~\kmps at $z = 1$ for the 1667 MHz main line). We used a single 6-hour track, with $\sim$4.7 hours on-source. J1939-6342 was observed to calibrate the absolute flux scale and bandpass, while J1347+1217 was used to correct for the time-varying complex gains. The target field was centred on RA = 14h29m35.2s, Dec = -00d28m36s.

We employed the \textsc{Oxkat} data processing pipeline \citep{2020ascl.soft09003H, 2024MNRAS.534...76H}. These scripts carry out radio frequency interference (RFI) excision, amplitude, bandpass, and cross-calibration. With user intervention and inspection, the scripts also perform both direction-independent and direction-dependent self-calibration and imaging.  \textsc{Oxkat} incorporates several radio astronomy packages, all described in \citet{2020ascl.soft09003H}.

The combination of the wide field of view ($>5$~deg$^2$ at 816~MHz) and lower observing frequency results in a target field that includes multiple bright continuum sources ($> 100$~mJy), leading to a dynamic-range-limited image using direction-independent self-calibration. To enhance the image quality, the strong bright sources were `peeled' from the target field visibilities using \textsc{CubiCal}. The peeling process involves applying calibration solutions to a phase-centred bright source prior to its visibility-plane subtraction from the target data set, followed by the next round of imaging. We removed two bright sources ($> 100$~mJy) from our dataset, which were identified through visual inspection. 

For spectral-line imaging, we split out a 300-channel subset of the full-resolution 32k dataset centred on the frequency range where the target source's redshifted 1667~MHz OH emission line is expected.  The continuum is subtracted from the calibrated visibilities using \textsc{Casa}'s \texttt{uvsub} and \texttt{uvcontsub} tasks, assuming a linear spectral model for the latter. Following this, spectral-line imaging was performed using \textsc{wsclean}, generating an OH emission cube. The imaging was performed using a \textsc{briggs} \citep{1995PhDT.......238B} weighting scheme with \textsc{robust} = 0.5, which was found to be a reasonable tradeoff of OH line signal-to-noise and angular resolution for our goals. The resultant circularised Point Spread Function (PSF) full-width half-maximum is 32.08~arcsec, while the achieved 16.6 kHz channel map median rms is $\sigma = 362~\mu$Jy\,beam$^{-1}$.

Finally, residual continuum emission was subtracted assuming a third-order polynomial using \textsc{casa} task \texttt{imcontsub}. 

There is no evidence in the individual channel maps (of a variety of {\sc robust} weightings), nor in the integrated moment 0 images that the OH emission is spatially resolved. We extract a spectrum using a aperture matched to the PSF FWHM dimensions noted above. The resultant observed OHM emission spectrum of the lensed system, HATLAS1429-0028 is shown in Fig.~\ref{fig:ohm_spectrum}, the features of which we discuss in greater detail following an explanation of the lens modelling.

%%%%%%%%%%
% Lens Model
%%%%%%%%%%%%

\section{Lens Model}\label{sec:lensing}

\begin{figure}
    \centering
	\includegraphics[width=0.4\textwidth]{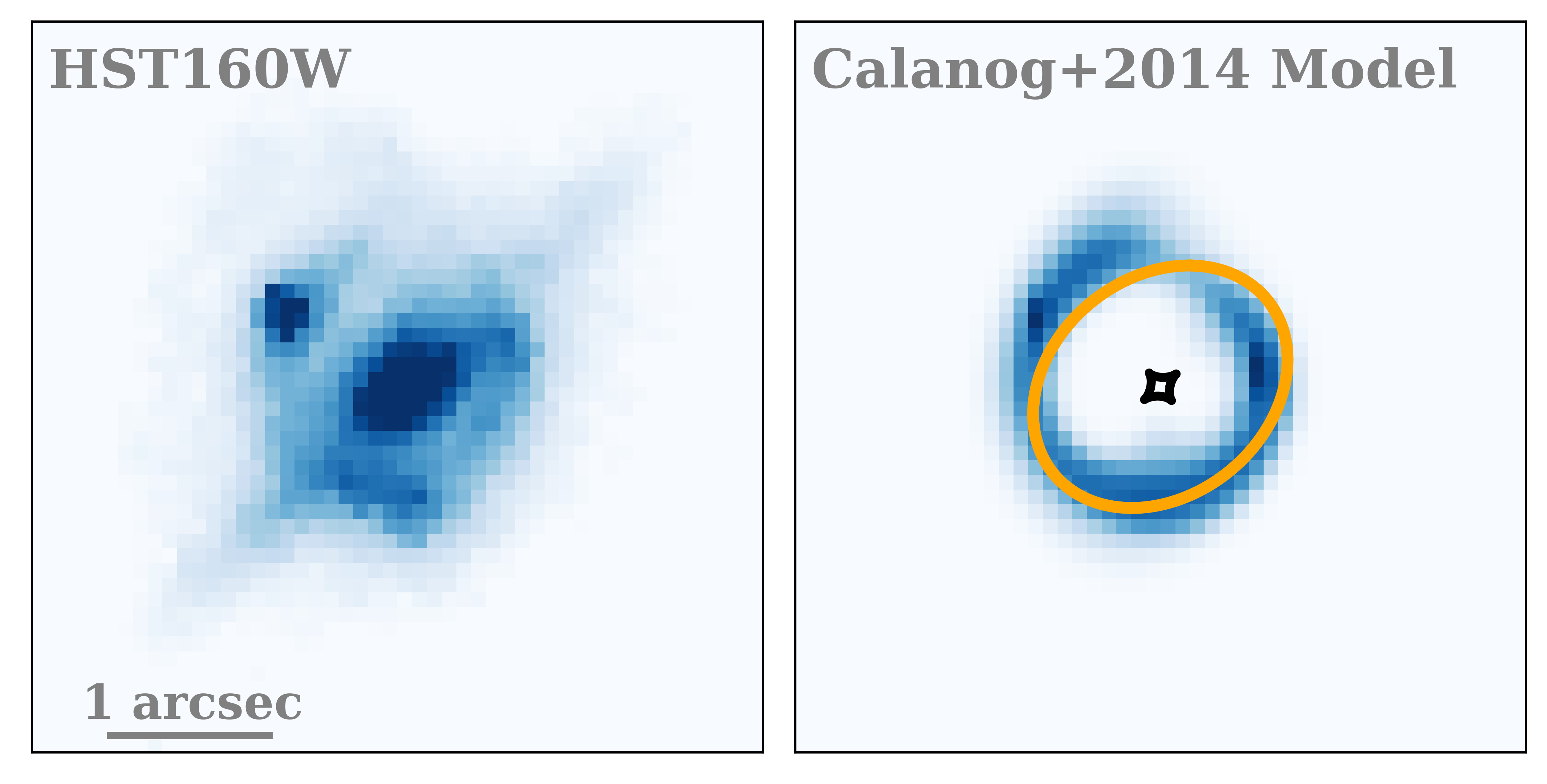}
    \caption[\obj Lens Model]{Left: {\sl Hubble Space Telescope} near-infrared (F160W) image of \obj, including the foreground disk lens. Right: \citet{Calanog2014} lens model of \obj, oriented with north at the top and east to the left. The orange and black lines denote the critical and caustic curves, respectively.} 
    \label{Fig:lensmodel}
\end{figure}

 We adopt the lens model derived in \citet{Calanog2014} who employ an interactive approach using {\sc galfit} and {\sc gravlens} \citep{Keeton01} to reconstruct the Keck near-infrared data. They assume a Single Isothermal Ellipsoid (SIE) mass-density profile and no external shear, with the foreground lens at a redshift of $z_{\rm l} = 0.218$ and a background source at $z_{\rm s} = 1.027$. The derived SIE model for \obj\ has an Einstein radius $\theta_{\rm E} = 0.738^{+0.002}_{-0.001}$~arcsec.  The SIE minor-to-major axis ratio (ellipticity) is $q = 0.792^{+0.005}_{-0.003}$, at a position angle of $\theta = -51.0^{+0.5}_{-0.4}\deg$ (east of north). The modelling suggests an offset between the lens centre and the brightest component of the foreground disk galaxy light, which they constrain to be $\delta x = 0.027^{+0.002}_{-0.002}$~arcsec, $\delta y = 0.044^{+0.002}_{-0.003}$~arcsec.

\citet{Calanog2014} use three source-plane components to model the near-infrared emission, noting that any fewer increases the reduced $\chi^2 = 2.6$ by a factor of $\sim2$ and making the case that this is a galaxy merger system. In Fig.~\ref{Fig:lensmodel}, we show their lens model, which we employ to constrain the magnification of the detected OH megamaser. We note that \citet{messias2014herschel} also model the NIR data and find a consistent model and derived magnifications. \citet{messias2014herschel} also model several molecular gas tracers with coarser spatial resolution to the NIR data.  Like \citet{Calanog2014}, they find strong evidence for complex source morphologies, which they also argue is likely a major merger system, consistent with the host galaxy properties of OHM megamasers. 

Constraining the magnification of the OHM is fundamentally limited by MeerKAT's $\sim$10~arcsec angular resolution at these frequencies and the resultant astrometric accuracy and precision, alongside our ignorance of the compactness of the OH emission. For these reasons, we adopt the \citet{Calanog2014} NIR model and only seek to make statements based on this macro model. 

To contextualise the modelling approach, we make the following observations. The OHM is likely to be associated with the merger activity, and therefore with one more of the NIR cores, and potentially between the two NIR nuclei, as argued by \citet{messias2014herschel}. This configuration where dense gas falls to the centre of the halo potential, between the stellar cores is seen in analogous systems  \citep[e.g. `The Antennae', NGC\,4038/4039, ][]{Hibbard2001}. In addition, the OHM spectrum is made up of two narrow components with FWHM of ($\lesssim 25$~km\,s$^{-1}$) alongside a broader component ($\sim300$~km\,s$^{-1}$), where the latter is typically interpreted as OH emission arising from a larger emission component \citep[e.g.][]{Glowacki,jarvis2023discovery}. Taken together, we argue the following approach offers an indicative view of what the magnification of the OHM emission (components) may be, within current spatial resolution and astrometric uncertainty limits. 

\begin{enumerate}
    \item We assume the OHM emission maximum size  $\theta_{\rm OH} \lesssim$1~arcsec, based on the PSF dimensions ($\sim 30$~arcsec), unresolved nature, and maximum SNR of the detection in individual channel maps. 
    \item The OH could be associated with either nucleus, between them, and/or in the larger scale diffuse component.
    \item To estimate the range of OHM magnification factors possible, we adopt all three NIR component positions listed in \citet{Calanog2014}, and explore a wide range of possible source-plane radii between 0.002 - 1 arcsec ($\sim 15-8000$~pc). The upper limit is set by the approximate astrometric precision of the OHM detection, while the lower limit is set by a combination of computational processing considerations and OHM size constraints. 
\end{enumerate}

\begin{figure}
    \centering
    \includegraphics[width=0.48\textwidth]{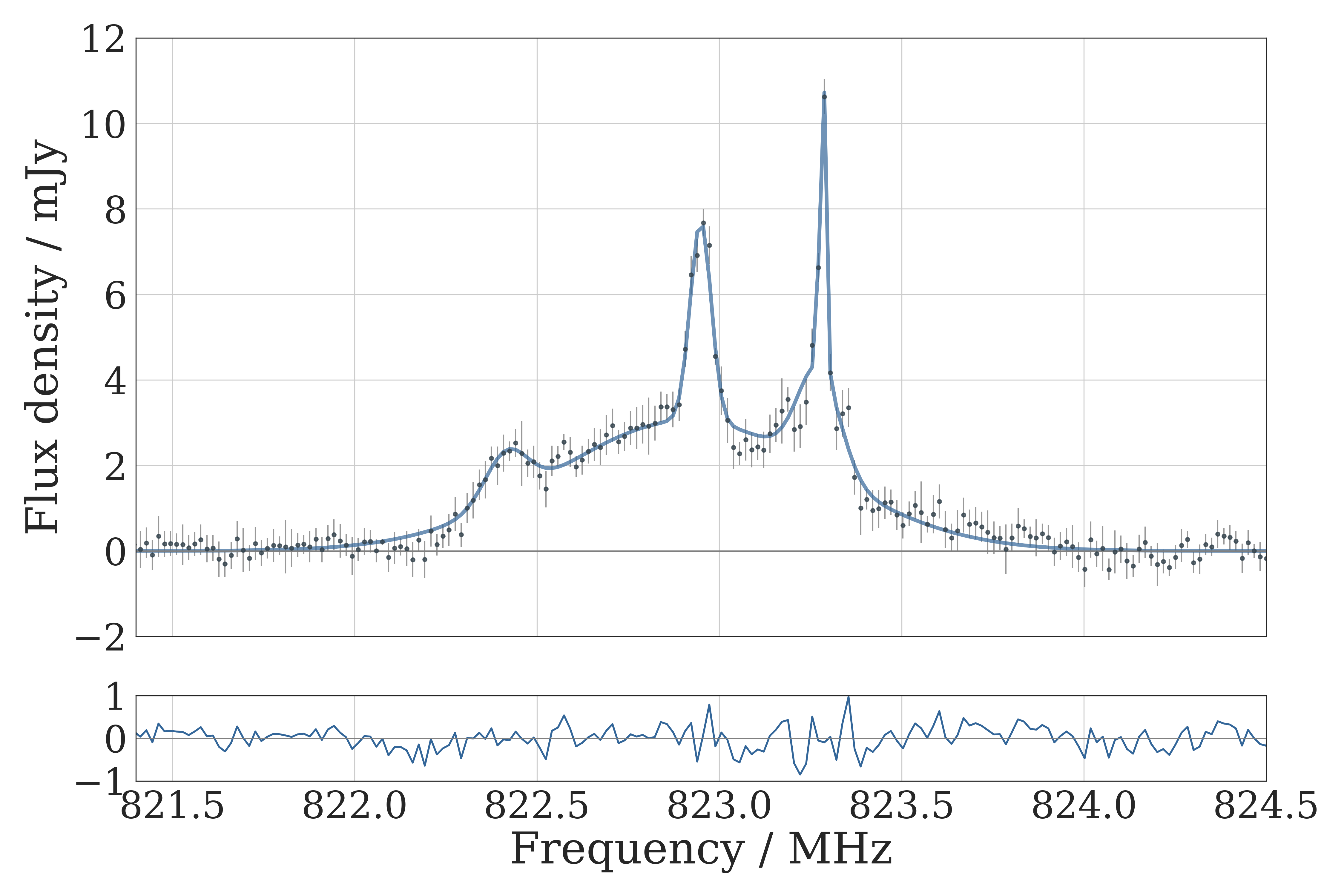}
    \caption{Observed OHM emission spectrum of the lensed system, HATLAS1429-0028. The 16.6~kHz resolution spectrum shows a complex emission profile with remarkably high integrated SNR. Bayesian model selection prefers the 5-Gaussian component model (blue curve). The residuals (data - model) are shown in the bottom panel. }
    \label{fig:ohm_spectrum}
\end{figure}

\begin{figure}%[hbt!]
\centering
	\includegraphics[width=0.47\textwidth]{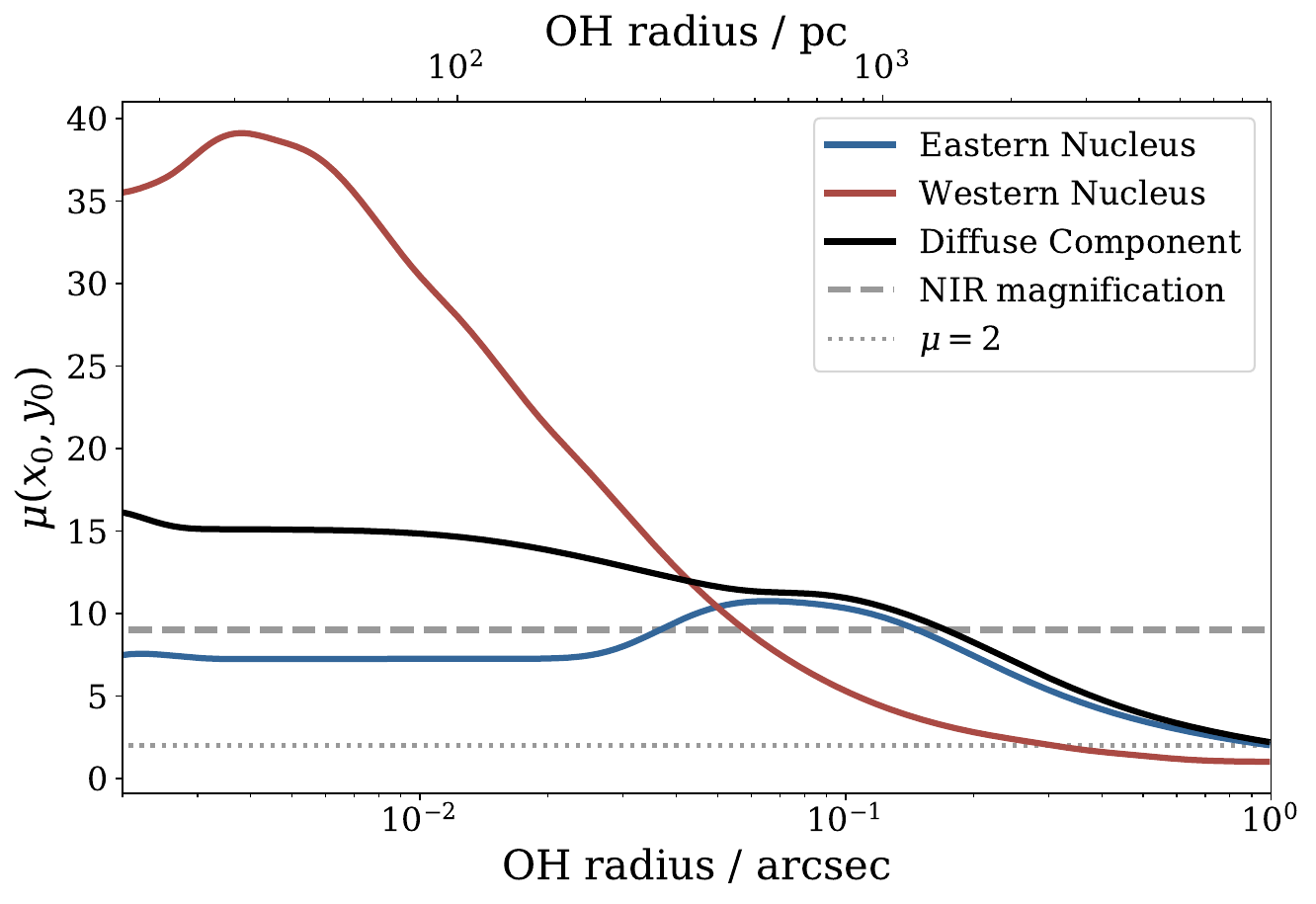}
    \caption[\obj Lens Model]{The relationship between magnification and radius for hypothetical OH maser components centred on the three NIR components identified in the \citet{Calanog2014} lens modelling with Keck {\sl K}$_{\rm s}$ Adaptive-Optics imaging data. We centre circularly symmetric Gaussian components on these three locations and compute the magnifications for each. The dashed horizontal line shows the total NIR magnification derived from \citet{Calanog2014}, which is consistent with the value derived in \citet{messias2014herschel}. See the main text for the full discussion. }
    \label{Fig:Magnification}
\end{figure}

\noindent We use the {\sc lenstronomy} package \citep{Birrer2018,Birrer2021} to carry out all necessary ray-tracing realisations in this work, utilising a supersampling parameter that ensures the source-plane radii are always sufficiently well-sampled. The result of this approach is an indicative range of magnifications of the three components which serve to illuminate more detail than the simple magnification $\mu \sim 9$ constraint we could otherwise adopt from high-resolution molecular gas observations and modelling carried out in \citet{Calanog2014} and \citet{messias2014herschel}. In Fig.~\ref{Fig:Magnification}, we show the magnification of each component as a function of radius, in both angular and spatial scales. There are two key conclusions to draw from this plot: (i) the Eastern Nucleus and the Diffuse Component reach maximum magnifications of $\sim10$ for source effective radii of $r_{\rm eff} \sim 0.1$ arcsec, similar to their best-fit source-plane radii presented in \citet{Calanog2014}. For sizes smaller than these, both the Eastern Nucleus and the Diffuse Component have a relatively constant magnification of $\sim9-15$, respectively. In strong contrast, the modelled magnification of the Western Nucleus, which lies much closer to the caustic, continues to increase with decreasing source size. For sizes consistent with that typically seen in OHMs ($10-100$~pc), the magnification continues to rise to $\sim40$, a factor of $\sim3-4$ larger than the other components.
\\\\
At face value, this provides a plausible explanation for some of the OHM integrated spectrum properties seen in Fig.~\ref{fig:ohm_spectrum}. First, the broad velocity component likely stems from a larger solid angle (and/or background continuum source)  that still undergoes a relatively large magnification factor of $\sim10$, consistent with the NIR magnification of $\mu_{\rm NIR} = 9.6^{+1.0}_{-0.3}$ \citep{Calanog2014}. Second, the compact nature of OH maser spots and close proximity of the Western nucleus to the caustic may result in a highly magnified $\mu \gtrsim 30$ masering region, explaining the high SNR, extremely narrow OH emission line seen in Fig.~\ref{fig:ohm_spectrum}. This is also likely linked to a compact background radio continuum source tracing an AGN, which would undergo a higher magnification, as Fig.~\ref{Fig:Magnification} demonstrates.

\section{Results and Discussion}

%%%%%%%%%%%%%%%%%%%%%%%%%%%%%%%%%%%%%%%%%%%%%
\subsection{OHM spectrum modelling}
\label{sec:H1429-BeyesTheorem-OHM-modelling}
%%%%%%%%%%%%%%%%%%%%%%%%%%%%%%%%%%%%%%%%%%%%%%

The observed OHM spectrum, extracted using the procedure discussed in Sec.~\ref{sec:Observations-Dataprocessing-calibration}, is shown in Fig.~\ref{fig:ohm_spectrum}, revealing a line profile composed of multiple components. Based on the complex OHM line profile and high achieved SNR, we opt for a model comprised of a sum of multiple Gaussians:

\begin{equation}
\label{GaussianModel}
    s(\nu) = \sum\limits_{i=0}^N A_i
    \exp\left(-\frac{(\nu - \beta_i)^2}{2\sigma_i^2}\right), 
\end{equation}

\noindent where $s(\nu)$ denotes the combined spectral line profile, and $\beta_i$, $\sigma_i$, $A_i$ are the mean, standard deviation, and peak flux density of each Gaussian component, respectively. The priors for $\beta_i$ are bounded uniform distributions, enforcing $\beta_0 < \beta_1 < \beta_2 < \beta_3 < \ldots < \beta_N$, thereby avoiding multimodal distributions and non-convergence. For the amplitude $A_i$, we set a prior of $A_i > 3\sigma$, where $\sigma = 0.362$~mJy beam$^{-1}$ in a 16.6 kHz channel.  If the total number of Gaussians is $N_{\rm g} \geqslant 4$, we relax this prior to $A_i > 1\sigma$ to facilitate convergence. The Gaussian width, defined by $\sigma_i$, has a minimum value of half a channel, i.e. $\sigma_i\geqslant$ 8.301~kHz. For $N_{\rm g} \geqslant 4$, we find this lower limit must be decreased to $\sigma_i \geqslant$ 1~kHz to accurately model the brightest OH spectral peak.

\noindent We employ nested sampling \citep{Skilling}, specifically the \textsc{pymultinest} package \citep{buchner2014x}, to constrain the model parameters described in Equation~\ref{GaussianModel} and calculate the Bayesian evidence, enabling robust model selection. Using the Jeffreys' scale \citep{jeffreys1998theory}, the Bayes factor only provides moderate evidence in support of the 6-Gaussian over the 5-Gaussian model. In contrast, the 5-Gaussian model is strongly favoured over models with four Gaussian components or less (i.e. Bayes Factors $> 3$). We therefore select the 5-Gaussian model for the remainder of our analysis. In Fig.~\ref{fig:ohm_spectrum}, we show the spatially unresolved OHM spectrum with the median posterior 5-Gaussian model overplotted. The properties of each component in this 5-Gaussian model are listed in Table~\ref{tab:H1429-ohm-properties}. We plot the residual spectrum in the lower panel of Fig.~\ref{fig:ohm_spectrum}.

\noindent The OHM spectrum and its model fit have several striking properties, which we briefly summarize below.
\begin{enumerate}
    \item The integrated OHM luminosity is the most apparently luminous known with $\log(\mu L_\textmd{{OH}}$ / $L_{\odot})$ = 5.51$\pm$0.67. If we assume the NIR-derived magnification factor for the OHM emission ($\mu \sim 10)$, then this remains amongst the most luminous with $\log(L_\textmd{{OH}}$ / $L_{\odot}) = 4.5$. 
    \item The brightest spectral line component corresponds to a very narrow component with full-width half-maximum (FWHM) of $\Delta V = 7.05 \pm 1.27$~\kmps ($\Delta \nu = 19 \pm 3$~kHz). Given that individual OH maser emission components can be highly compact ($\lesssim 10$~pc), they can, therefore, in principle, undergo extremely high lensing magnifications $\mu \gg 100$.
    \item The second-highest OH peak does not have a central frequency corresponding to the 1665~MHz line, as one may naively expect from the OHM line profile. The measured frequency offset between the two brightest peaks is $ \Delta \nu \sim0.34$~MHz, not  $\Delta \nu = 1.9572/(1+z) = 0.9656$~MHz, as one would measure if they corresponded to the 1665 and 1667 lines at the same observed-frame velocity. 
    \item Despite the high OHM luminosity, the measured width of the broad component ($\Delta V = 315 \pm 10$~\kmps) is lower than typically seen in low-redshift OHM systems 
    \citep[e.g.,][]{pihlstrom2001evn}, as well as the two other MeerKAT-discovered high-redshift systems at $z = 0.5$ and $z = 0.7$ \citep{Glowacki,jarvis2023discovery}. 
    \item The apparent OH luminosity is consistent with the predicted OH luminosity of log($\mu L_\textmd{{OH}}$ / $L_{\odot}$) = 5.51, derived from the $L_\textmd{{FIR}}$-$L_\textmd{{OH}}$ correlation \citep{zhang2024fashi}. At face value, this suggests that the total far-infrared and OH magnification factors are likely comparable ($\mu \sim 10$), and that an extreme magnification maser component (e.g. $\mu \gtrsim 100$) is unlikely to dominate the apparent OHM luminosity.
\end{enumerate}

Disentangling this blended spectrum clearly requires substantially higher angular resolution ($\lesssim 0.1$~arcsec). However, comparison with the cold molecular gas spectra \citep{messias2014herschel}, and the newly-discovered \hi absorption, described below, can provide a more holistic picture of the OHM system. 

%%%%%%%%%%
%% FIGURE %%
%%%%%%%%%%
\begin{figure}
\centering
\includegraphics[width=0.3\textwidth]{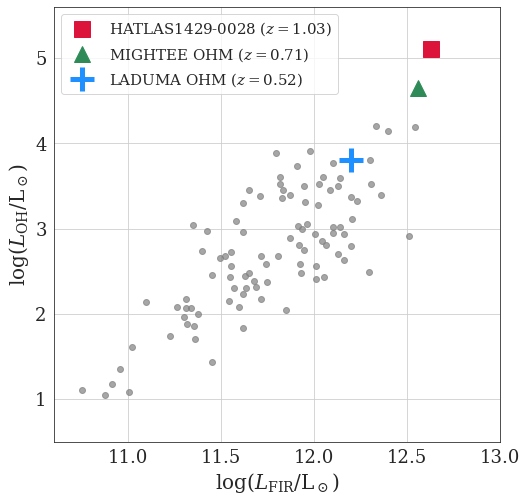}
\caption{Far-infrared and OH luminosity correlation \citep[data points from][]{Glowacki,jarvis2023discovery,zhang2024fashi}. The three MeerKAT-discovered OHMs (blue, green, red) have doubled, tripled, and quadrupled the previous OHM redshift record held for almost two decades, respectively. The MIGHTEE (green) and highest redshift source (red) reported in this work are both strongly lensed, suggesting untargeted OHM searches as a way of identifying lensed systems \citep[see, ][]{Button2024}.}
\label{Fig:H1429-lum_correlation}
\end{figure}

\begin{table}
    \caption{Median posterior parameters of the five-component Gaussian OH spectral line model, where each component's mean, FWHM, integrated flux, integrated apparent luminosity, and 1667-to-1665 line ratio is denoted by $\nu_{\text{obs}}$, $\Delta v_{\text{rest}}$, $F_{\text{line}}$ , log$(\mu L_{\text{OH}})$, and $R^{1667}_{1665}$, respectively.}
    \label{tab:H1429-ohm-properties}
    \centering
    \scriptsize
    \begin{tabular}{c  c c c c c}
        \hline
        \#  & $\nu_{\text{obs}}$ & $\Delta v_{\text{rest}}$ & $F_{\text{line}}$ & log$(\mu L_{\text{OH}})$  & $R^{1667}_{1665}$ \\
                 & MHz                      & km~s$^{-1}$            & Jy~km~s$^{-1}$  & $L_{\odot}$           &           \\
            \hline \hline
        1  & $822.42 \pm 0.012$ & $50.3 \pm 11.2$ & $0.14 \pm 0.04$ & $3.8 \pm 1.0$ & - \\
        2  & $822.91 \pm 0.015$ & $314.8 \pm 9.5$ & $2.41 \pm 0.10$ & $5.0 \pm 0.2$ & $17.2 \pm 5.0$ \\
        3  & $822.95 \pm 0.002$ & $24.5 \pm 1.6$ & $0.29 \pm 0.05$ & $4.0 \pm 0.3$ & $2.1 \pm 0.7$ \\
        4  & $823.27 \pm 0.006$ & $49.2 \pm 5.8$ & $0.30 \pm 0.05$ & $4.1 \pm 0.7$ & $2.1 \pm 0.6$ \\
        5  & $823.28 \pm 0.001$ & $7.1 \pm 1.3$ & $0.13 \pm 0.03$ & $3.7 \pm 0.9$ & $0.9 \pm 0.3$ \\
        \hline
    \end{tabular}
\end{table}

\subsection{\hi absorption}

\begin{figure}
	\includegraphics[width=0.48\textwidth]{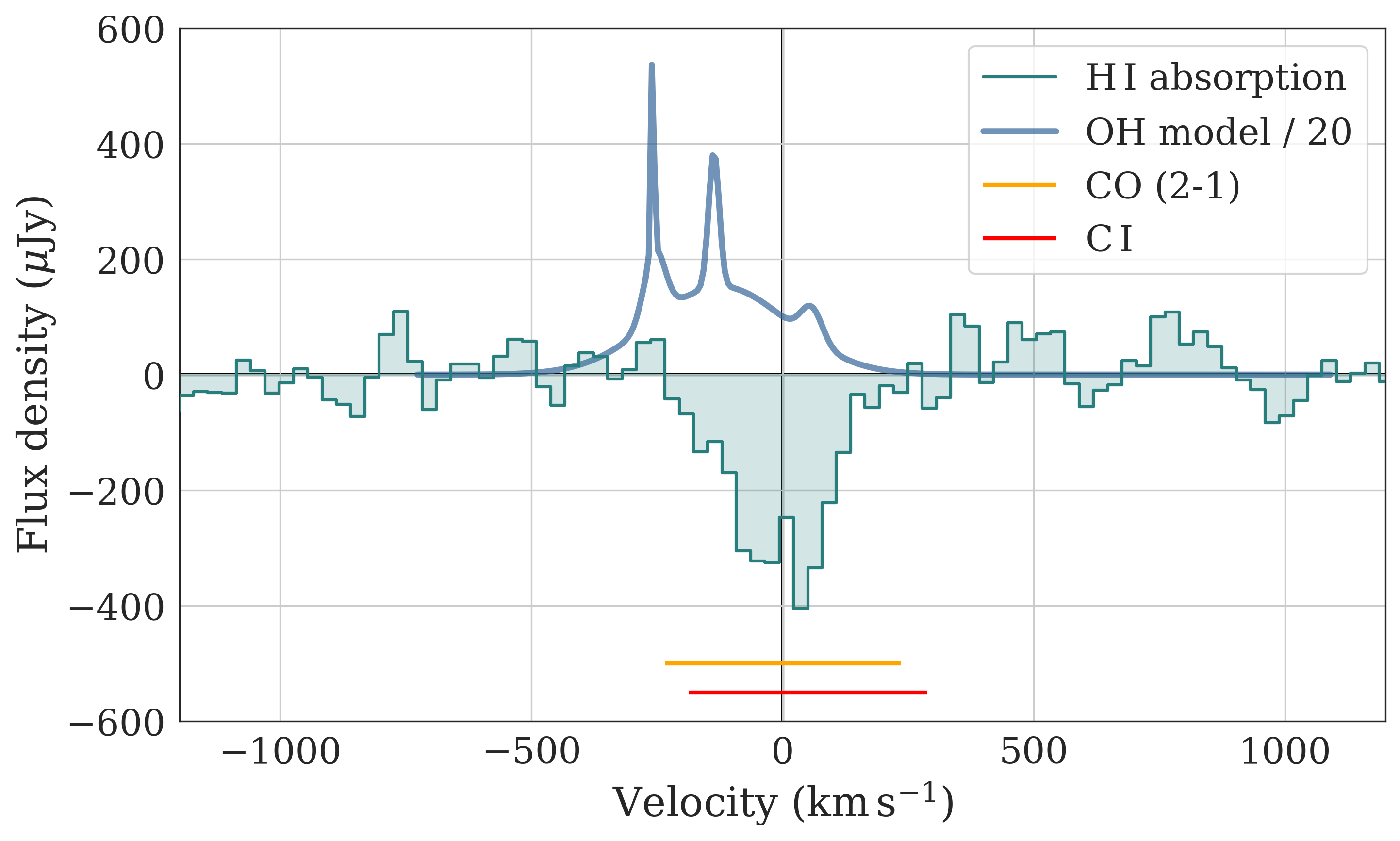}
    \caption[\hi absorption spectrum]{The \hi absorption spectrum (light blue), which is centred on the systemic redshift ($V=0$~km\,s$^{-1}$), as derived from optical and low-J CO lines. The OH emission spectrum (dark blue), scaled by 0.04 for comparison, appears blue-shifted, potentially indicating a molecular outflow. }
    \label{Fig:HI_absorption}
\end{figure}

The \obj~MeerKAT \hi absorption spectrum is shown in Fig.~\ref{Fig:HI_absorption}. Bayesian model selection, applied with the same approach as the OHM spectrum, favours a two-Gaussian-component model. The median posterior centroid velocities of these two \hi components are symmetric about the systemic redshift ($V_{\rm c} = -36.3\pm2.0$ and $18.1\pm3.7$~km~s$^{-1}$), with very similar FWHM = 38.2$\pm$3.6 and 44.1$\pm$5.9~km~s$^{-1}$; and peak flux densities of $s_{\rm HI}^{\rm peak} =  -0.58 \pm 0.03$ and -0.44$\pm$0.05~mJy. We estimate the \hi column densities of the fitted components to be $N_{\rm {H\,\textsc{i}\xspace}} = 1.21 \times 10^{21}$ and $1.09 \times 10^{21}~\rm cm^{-2}$, assuming a spin temperature $T_s = 100$~K and the covering factor of the background continuum source $f_c = 1$. 

The OHM spectrum median posterior model (scaled for comparison) is overlaid on the \hi spectrum, in the rest frame, for comparison. Additionally, the velocity positions and corresponding FWHM of the CO and \ci emission lines are indicated, as adopted from \citet{messias2014herschel}. The mean of the two \hi absorption components is consistent with that measured from optical spectroscopy, as well as the low-$J$ CO and \ci emission tracing colder molecular gas. The \hi velocity width is, however, much narrower than the CO and \ci emission FWHM, suggesting it may be further out in the galaxy and/or is associated with a significantly smaller gas component relative to the entire system. The OHM emission, particularly the two brightest peaks, are clearly blueshifted relative to the colder neutral and molecular gas, suggesting that the OHM is potentially tracing a warm gas molecular outflow. 

\subsection{A molecular outflow in \obj?} \label{sec:outflow}

Comparison with other OHMs (see Fig.~\ref{Fig:H1429-lum_correlation}) distinguishes \obj~as the most apparently luminous and distant OHM known, placing it on the borderline of being classified as an OH GigaMaser (OHGM). One, therefore, asks the question of whether an AGN is contributing to the maser pumping mechanism. To explore this, we calculate the far-infrared to radio continuum luminosity ratio, $q_{\rm{FIR}}$. We estimate $q_{\rm{FIR}}$ = 2.2$\pm$0.2, which is within 2$\sigma$ of the median value reported in \citet{ivison2010far} for star-forming galaxies. This suggests that the radio emission is potentially powered by both star formation and radio AGN activity, however, we cannot exclude contamination from the foreground lens or significant differential magnification between AGN and star formation components. 

The measured dust temperature of $T_{\rm dust}~\sim$~40.7~K \citep{ma2018sofia}, is broadly consistent with known OHMs at higher redshifts \citep{Glowacki,jarvis2023discovery}, as well as with theoretical predictions that a dust temperature of at least 45~K is necessary for OH maser activity \citep{lockett2008effect,willett2011mid}. However, we note that this average dust temperature may be lower than dust in the nuclear region where the OHM emission is expected to be, particularly if an AGN and nuclear starburst are present. 

At face value, the relative velocities of the integrated OH, H\,\textsc{i}, CO (2-1), and C\,\textsc{i} line profiles suggest the presence of a warm molecular gas outflow, potentially driven by AGN activity. However, the complexity of the OH spectrum indicates the presence of at least two dominant maser components, separated by $\sim$120~km\,s$^{-1}$. One of the simplest interpretations would be that these components arise from two merging nuclei with a modest systemic velocity offset, as observed in systems such as Arp~220, where VLBI observations reveal OH megamaser emission associated with both nuclei \citep{diamond1989vlbi,Baan2023}. In such cases, the superposition of emission from multiple nuclei (and the associated disrupted gas) can also produce an apparently broad component in the integrated spectrum \citep{Pihlstrom2004}. Alternative scenarios are also viable. For example, \citet{Tegegne2025} report spatially resolved OH and radio continuum observations of IRAS~15250+3609, revealing a clear spatial offset between the OH and continuum peaks. Similar to \obj, its integrated spectrum exhibits two prominent 1667~MHz components, which they interpret as low- and high-gain maser amplification associated with an AGN and shock-heated gas from a foreground infalling dwarf galaxy, respectively, consistent with the narrow and broad velocity widths of the two components. While both spatially resolved scenarios are plausible for \obj, the interpretation is further complicated by the presence of strong gravitational lensing, particularly the effects of differential magnification, as outlined in Sec.~\ref{sec:lensing}. Discriminating between these possibilities remains challenging with the current angular resolution, underscoring the need for future high-angular-resolution OH observations to better constrain the lens model and robustly distinguish between the scenarios outlined here.

\section{Summary}\label{sec:conclusion}

We have observed OH 18-cm emission and \hi 21-cm absorption in the gravitational lens system \obj~at $z$ = 1.027, which is the record cosmological distance for an OHM known to date. The OH spectrum reveals four frequency-resolved components, with at least two of these likely originating from different regions of the lens and experiencing significant magnification ($\mu > 10$). The total (apparent) OH luminosity is measured as log($\mu L_{\rm OH}~/~L_{\odot}$) = 5.51~$\pm$~0.67. Assuming a NIR magnification of $\mu \sim 10$ places \obj in the gigamaser class and implies that it is the most luminous known system to date. This discovery highlights MeerKAT's potential to investigate high-redshift OHMs, enhancing our understanding of thereof and offering valuable tracers for exploring different aspects of galaxy outflows and merging activity. Additionally, similar to \hi studies, gravitational lensing offers a promising method for studying high-redshift OHMs using the MeerKAT telescope. However, a long-baseline ($\gtrsim 40$~km) component of the SKA-Mid will be essential to fully realise the rich scientific yield from these systems, particularly given the large number of lensed OHM systems expected in the SKA era \citep{Button2024}.

\section*{Acknowledgements}
We thank the anonymous referee for their helpful suggestions. We thank Jae Calanog for helpful discussions on the lens modelling of  \obj. TEM, RPD, TB, and IH acknowledge support from the South African Radio Astronomy Observatory, which is a facility of the National Research Foundation (NRF), an agency of the Department of Science, Technology and Innovation (DSTI). RPD's research is funded by the South African Research Chairs Initiative of the DSTI/NRF. DO is a recipient of an Australian Research Council Future Fellowship (FT190100083) funded by the Australian Government. The MeerKAT telescope is operated by the South African Radio Astronomy Observatory, which is a facility of the National Research Foundation, an agency of the Department of Science and Innovation. We acknowledge the use of the ilifu cloud computing facility – www.ilifu.ac.za, a partnership between the University of Cape Town, the University of the Western Cape, Stellenbosch University, Sol Plaatje University and the Cape Peninsula University of Technology. The ilifu facility is supported by contributions from the Inter-University Institute for Data Intensive Astronomy (IDIA – a partnership between the University of Cape Town, the University of Pretoria and the University of the Western Cape), the Computational Biology division at UCT and the Data Intensive Research Initiative of South Africa (DIRISA). This work made use of the CARTA (Cube Analysis and Rendering Tool for Astronomy) software (DOI: 10.5281/zenodo.3377984 – \url{https://cartavis.github.io}). Based on observations made with the NASA/ESA Hubble Space Telescope, and obtained from the Hubble Legacy Archive, which is a collaboration between the Space Telescope Science Institute (STScI/NASA), the Space Telescope European Coordinating Facility (ST-ECF/ESA) and the Canadian Astronomy Data Centre (CADC/NRC/CSA). IH acknowledges support of the STFC consolidated grant [ST/S000488/1] and [ST/W000903/1], and from a UKRI Frontiers Research Grant [EP/X026639/1]. IH acknowledges support from Breakthrough Listen. Breakthrough Listen is managed by the Breakthrough Initiatives, sponsored by the Breakthrough Prize Foundation.

%%%%%%%%%%%%%%%%%%%%%%%%%%%%%%%%%%%%%%%%%%%%%%%%%%
\section*{Data Availability}

The data relevant to this Letter are publicly available in MeerKAT archive \url{ https://archive.sarao.ac.za}.



\bibliographystyle{mnras}
\bibliography{refs} % if your bibtex file is called example.bib

%%%%%%%%%%%%%%%%%%%%%%%%%%%%%%%%%%%%%%%%%%%%%%%%%%

%%%%%%%%%%%%%%%%%%%%%%%%%%%%%%%%%%%%%%%%%%%%%%%%%%

% Don't change these lines
\bsp	% typesetting comment
\label{lastpage}
\end{document}